\documentclass[12pt]{iopart}
\usepackage{epsfig,graphics}
\usepackage{amssymb}
\newcommand{\pT}{\ensuremath{p_{T}}}
\newcommand{\EBGest}{\ensuremath{\hat{E}_{T}^{BG}}}
\newcommand{\EBGtrue}{\ensuremath{E_{T}^{BG}}}
\begin{document}


\title{A Cone Jet-Finding Algorithm for Heavy-Ion Collisions at LHC Energies}

%


\author{S-L Blyth$^{1,2}$, M J Horner$^{1,2}$, T Awes$^3$, T Cormier$^4$, H Gray$^{1,2}$, J L Klay$^5$, S R Klein$^1$, M van Leeuwen$^1$, A Morsch$^6$, G Odyniec$^1$ and A Pavlinov$^4$}
\address{$^1$ Lawrence Berkeley National Laboratory, Berkeley, California 94720, USA}
\address{$^2$ UCT-CERN Research Centre, University of Cape Town, Rondebosch 7700, South Africa} 
\address{$^3$ Oak Ridge National Laboratory, Oak Ridge, Tennessee 37831, USA}
\address{$^4$ Wayne State University, Detroit, Michigan 48202, USA}
\address{$^5$ Lawrence Livermore National Laboratory, Livermore, California 94550, USA}
\address{$^6$ CERN, CH-1211 Geneva 23, Switzerland}

\begin{abstract}
Standard jet finding techniques used in elementary particle collisions have
not been successful in the high track density of heavy-ion
collisions. This paper describes a modified cone-type jet finding
algorithm developed for the complex environment of heavy-ion
collisions.  The primary modification to the algorithm is the
evaluation and subtraction of the large background energy, arising from
uncorrelated soft hadrons, in each collision. A detailed analysis of
the background energy and its event-by-event fluctuations has been
performed on simulated data, and a method developed to estimate the
background energy inside the jet cone from the measured energy
outside the cone on an event-by-event basis. The algorithm has been
tested using Monte-Carlo simulations of Pb+Pb collisions at
$\sqrt{s}=5.5$~TeV for the ALICE detector at the LHC.  The
algorithm can reconstruct jets with a transverse energy of 50 GeV and above with an energy
resolution of $\sim30\%$.
\end{abstract}

\pacs{25.75.Nq 13.87.Fh}

\section{Introduction}

Jet-finding techniques, a well-established tool for $p+p$,
$e^{+}+e^{-}$ and $e+p$ collisions~\cite{SoperEllisReview}, are not directly
applicable in heavy-ion (HI) collisions due to the overwhelming 
combinatorial backgrounds from high multiplicity underlying events. For central Pb+Pb collisions, nearly
all ~400 nucleons participate, leading to a high multiplicity of particles produced in simultaneous
nucleon-nucleon collisions. In conventional jet-finding algorithms, this background energy will
be swept up into the jet-cone, and strongly distort the reconstructed jet. This problem is also present,
but to a much lesser extent, in p+p collisions at the LHC due to multiple p+p interactions as a
result of the planned high luminosity, and due the high probability for multiple-parton interactions within
a single p+p interaction.

Experience from RHIC (Relativistic Heavy-Ion Collider) has shown
that high transverse momentum, \pT{}, phenomena are promising tools to investigate the
hot and dense medium produced in heavy-ion collisions. The most
striking results from the first five years of RHIC operation are
centered around the observation that hadron production at high \pT{}
is strongly suppressed in central heavy-ion collisions at $\sqrt{s_{NN}}=200$~GeV.  High-\pT{} hadrons are
dominantly produced by the fragmentation of partons from hard
scatterings in the initial state. The suppression of particle
production is likely due to energy loss by these partons in the hot
and dense matter created in the collision (possibly a Quark Gluon
Plasma)~\cite{GULPLUM,XNW1,XNW2,XNW3}. Reconstructing the remnants of
these hard partons, jets, can serve as a probe of the produced medium.

\begin{figure}[!hbt]
  \center
  \includegraphics[scale=0.4]{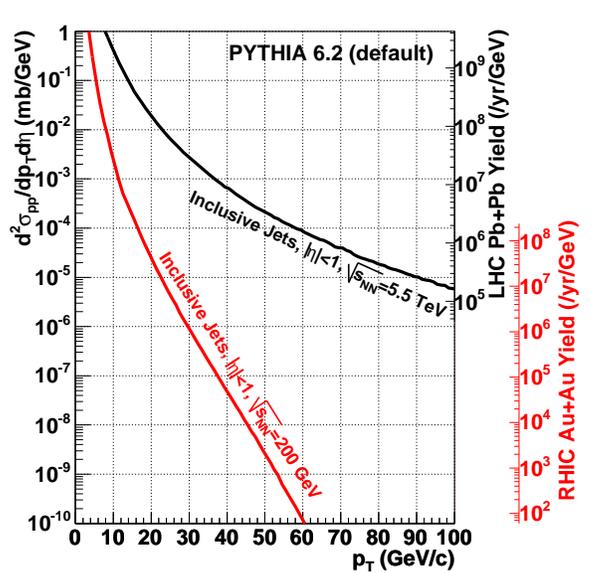}  
  \caption{\label{fig:edep_jets}Differential cross-section for
  inclusive jets within $|\eta|<1$ in $p+p$ collisions at RHIC and LHC
  from default {\sc pythia} 6.2 (left axis).  The annual yields shown on the right axes are for
  minimum bias Pb+Pb (Au+Au) collisions assuming 10$^6$ s (10$^{7}$ s)
  running time and 0.5 mb$^{-1}$ s$^{-1}$ (5.0 mb$^{-1}$ s$^{-1}$) 
  luminosity at LHC (RHIC).} 
\end{figure}

Measurements of parton energy loss at RHIC have been limited to
leading particle analysis and hadron correlations. Full jet
reconstruction and measurement of jet energy has not been possible due to
the large and highly fluctuating background of uncorrelated hadrons
of the underlying event and the typical low jet energies.  However, a striking improvement in jet reconstruction
capability is expected in heavy-ion collisions at $\sqrt{s_{NN}}=5.5$~TeV at the LHC (Large Hadron Collider).  An extrapolation of the charged
particle rapidity density based 
on Super Proton Synchrotron (17 GeV) and RHIC (130 and 200 GeV) measurements suggests an additional
factor of about 4 increase from RHIC top energy to LHC. The growth of the cross-section for hard processes is, however, 
much more dramatic.  Fig.~\ref{fig:edep_jets} shows the
differential cross-section for inclusive jets within a pseudorapidity range of $|\eta|<1$ in
$p+p$ collisions at RHIC and LHC energies from NLO pQCD as calculated by
{\sc pythia} 6.2~\cite{Pythia} (left axis) and
the expected annual yields in minimum bias Au+Au and Pb+Pb collisions
(right axis).  A substantial enhancement in the jet cross-section is already
seen at relatively low \pT{} ($\sim$20 GeV).  The kinematic reach for
jet measurements is thus much larger at LHC than at RHIC, allowing the
reconstruction of high energy jets above the uncorrelated background
on an event-by-event basis.

Theoretical studies of partonic energy loss in a quark-gluon plasma
predict that jets with intermediate transverse energies (50 GeV $\lesssim E_{T}
\lesssim$ 100 GeV) may provide the best probe of the highly excited
nuclear medium~\cite{CASUAW,VitevGyu,WangWang}.  Partons in this
energy range are expected to suffer 
the greatest relative energy loss and should therefore be more
useful for characterizing the properties of the medium than those at
asymptotically high energies.  Moreover, intermediate
energy jets at LHC are closer to the highest energies probed at RHIC,
making direct comparisons feasible. Thus, the emphasis of this
analysis is on the reconstruction of jets in the transverse energy range from 50 to 100 GeV.

A first attempt to identify and reconstruct jets on an event-by-event
basis in large background Pb+Pb collisions at the top LHC energy (5.5
TeV) is presented in this paper. A cone-type jet-finding algorithm has been 
adapted from the UA1 experiment~\cite{UA11} and further developed to
account for the heavy-ion background.  The choice of various
algorithm parameters is discussed and their influence on the jet
energy resolution is discussed. Two different methods for determining the large fluctuating background, the
most challenging problem for jet-finding, are evaluated.

The presented simulations are based on the ALICE detector, using
charged particle tracking for hadrons and an electromagnetic
calorimeter for photon and electron detection. It has been
shown previously that jets can be accurately reconstructed using
this combination of detectors in $e^{+}+e^{-}$ collisions at LEP (Large Electron Positron Collider)
~\cite{LEP} and $p+\bar{p}$ collisions at the Tevatron ~\cite{CDF}.
This paper demonstrates that it can also be successful in the analysis of more complex
heavy-ion collisions.

\section{The Heavy-Ion Jet Algorithm ({\sc hija}) and Simulations}

\subsection{Description of {\sc hija}}

This approach is based on a cone-type algorithm, developed by the UA1
collaboration~\cite{UA11}, where the jet is defined as a group of
particles in a cone of fixed radius in azimuth- ($\phi$) and
pseudorapidity- ($\eta$) space. The algorithm refinements evaluated by
the Tevatron Run II Jet Physics Group~\cite{RunII}, such as seedless
cones, splitting/merging corrections and $k_{T}$
algorithms~\cite{KT_algo}, are not considered here but their feasibility could
also be investigated for heavy-ion collisions.

The input to the algorithm is an energy grid in ($\eta$,$\phi$) filled
by a combination of transverse energy ($E_{T}^{Cell} =
E\sin\theta_{T}$, where $E$ is the total energy of the calorimeter
cell and $\theta_T$ is the polar angle of the cell) measured by the
electromagnetic calorimeter and charged track transverse momentum
($p_{T}$) information from the tracking system. The grid covers the
same fiducial area as the calorimeter, and each grid cell
corresponds in size and position to a calorimeter cell
($\eta\times\phi=0.014\times0.014$ in ALICE~\cite{AliceProp}).  In order to
reduce the contributions from uncorrelated background particles, only
charged hadrons with \pT{} above a threshold $p_{T}^{cut}$ were used
in the analysis.

Neutral energy is measured only in the calorimeter while charged
hadronic energy is registered in both the tracking detectors and in
the calorimeter. To correct for the double counting of hadronic energy, the
estimated energy deposited by charged hadrons in the calorimeter is
subtracted on a track-by-track basis using a parameterization of the
average simulated energy deposition of charged pions in the
calorimeter as a function of $\eta$ and $p_{T}$, $\langle
E_{HC}(\eta, p_{T})\rangle$.

The algorithm consists of the following steps:

\begin{enumerate}
\item Initialize the estimated background level per grid cell
  \EBGest{} to the average over all grid cells.

\item Sort cells in decreasing cell energy, $E_{T}^{i}$.

\item For at least 2 iterations, and until the change in
\EBGest{} between most recent successive iterations is smaller than a
set threshold
\begin{enumerate}
\item Clear the jets list
\item Flag all cells as outside a jet
\item Execute the jet-finding loop for each cell, starting with the largest:
\begin{enumerate}
\item \label{jetseed} If $E_{T}^{i}-\EBGest > E_{T}^{seed}$, where $E_{T}^{seed}$ is a
  chosen threshold cell energy, and the cell is flagged as not in a jet,
  treat it as a jet seed candidate:
\begin{enumerate}
    \item Set jet centroid ($\eta^{C}$, $\phi^{C}$) to
    the co-ordinates of the
    jet seed cell $\eta_{i}$, $\phi_{i}$.
    \item Using all cells within
      $\sqrt{({\eta}^i - {\eta}^C)^2+({\phi}^i-{\phi}^C)^2} <
      R$ of the initial centroid, calculate the new energy-weighted ($E_{T}^{i}-\EBGest$)
      centroid. Set the new energy-weighted centroid to be the new
      initial centroid.  Repeat centroid calculation iterations until the centroid
      does not shift by more than one cell in subsequent iterations.
    \item Store centroid as jet candidate and flag all cells within $R$
    of centroid as inside a jet.
  \end{enumerate}
  \end{enumerate}
  \item Re-calculate the estimated background energy \EBGest{} using the
    calculation described in Section \ref{sect:bg_est}.
    \item For each jet candidate, calculate the energy by summing the energies
    of the cells in the cone and subtracting the background. If the
    jet energy is greater than $E_{T}^{cone}$, the minimum allowed cone
    energy, a jet is found.
\end{enumerate}
\end{enumerate}

The main algorithm parameters and their purposes are listed in Table \ref{tab:Param1}.

\begin{table}[!hbt]
  \center \small
  \begin{tabular}{c|l}\hline \hline
    \textbf{Parameter}
    &
    \begin{tabular}{c}
    \textbf{Description}
    \end{tabular}
    \\
    \hline
    $E_{T}^{seed}$
    &
    \begin{tabular}{l}
      Minimum jet seed energy (after background subtraction)
    \end{tabular}
    \\
    $R$
    &
    \begin{tabular}{l}
      Radius of jet cone
    \end{tabular}
    \\
    \begin{tabular}{l}
    $E_{T}^{cone}$\\ $\ $
    \end{tabular}
    &
    \begin{tabular}{l}
      Minimum jet cone energy (after background subtraction)
    \end{tabular}
    \\
    $p_{T}^{cut}$
    &
    \begin{tabular}{l}
      Minimum track $p_{T}$
    \end{tabular}
    \\
    \hline \hline
\end{tabular}
\protect\caption{Main parameters used in {\sc hija}. }
\protect\label{tab:Param1}
\end{table}

\subsection{Description of Detector and Simulated Events}
\label{sec:simulation}
Simulations of Pb+Pb collisions were performed for the ALICE
experimental set-up using the ALICE
software framework, AliRoot ~\cite{AliRoot}.

The ALICE tracking detector response was approximated using a gaussian
smearing (with $\sigma$=1$\%$) of the track momentum $p$ and
a conservative tracking efficiency of $90\%$ (the presently
anticipated ALICE tracking efficiency is 98$\%$ ~\cite{AliceTDR}). The
ALICE electromagnetic calorimeter\footnote{ The final design which is still under evaluation. However, the changes under 
consideration are unlikely to affect jet reconstruction.}, a sampling calorimeter composed of
25 layers of 5mm $\times$ 5mm Pb-scintillator, was simulated
using {\sc geant}~3.21. The intrinsic energy resolution for photons
with energies from 25 to 200 GeV for this device was estimated from simulation to be
$\sigma(E)/E \sim 15\%/\sqrt{E}$~\cite{AliceProp}. The calorimeter was
simulated with a fiducial acceptance of $|\eta|<0.7, \pi/3<\phi<\pi$
and a granularity of 13~824 cells ($96 (\eta) \times 144 (\phi)$).

A sample of heavy-ion events with calibrated high energy jets was constructed by combining the output from two Monte
Carlo event generators.  Jet events were generated using {\sc pythia}
6.2 ~\cite{Pythia} and these were combined with high-multiplicity
Pb+Pb `background' events generated by {\sc hijing} 1.36
~\cite{Hijing}. To define the input jet energy scale in {\sc pythia}
events, the {\sc pycell} algorithm\footnote{{\sc pycell} is the
internal {\sc pythia} cone algorithm with $R=1$ which uses all
simulated particles to reconstruct jets.} was used.

Calibration samples of $E_T$=50 and 100 GeV ($\pm$5GeV)
jets were generated using {\sc pythia}. The jet directions were restricted in
pseudorapidity ($|\eta| < 0.3$) and in azimuthal angle (more than 0.26
radians from the edge of the calorimeter) to reduce the effect of
the acceptance edges on the energy reconstruction. The jet energy and
direction selection criteria were based on the output from {\sc pycell}. The
background event sample was composed of central
{\sc hijing} events (impact parameter\footnote{In
heavy-ion collisions, the impact parameter $b$ is defined as the
distance of closest approach between the centres of the colliding
nuclei. The most central collisions have the smallest impact
parameter.} $b < 5$~fm for the $10\%$ most central collisions). The
{\sc hijing} parameters were tuned for LHC energies according to
~\cite{YellowReport}. The charged particle rapidity density in these
events is approximately 4000 at mid-rapidity. This is 
likely to be an overestimate of the uncorrelated
background~\cite{ALICEPPR} at LHC.


\section{Background Energy Estimation and Choice of Algorithm Parameters}\label{AlgoOpt}

Parton energy loss effects are expected to be most visible in jets with
$E_T \lesssim$ 100 GeV ~\cite{CASUAW,VitevGyu,WangWang}.
The algorithm parameters were chosen to optimize the
energy resolution for jets with $E_T$= 50 GeV.

\subsection{Background energy estimation}
\label{sect:bg_est}
In this section simulated $p+p$ and Pb+Pb events are used to compare reconstructed jet
energies to the background level and optimise the background
estimation. 

\begin{figure}[!hbt]
  \center
  \includegraphics[scale=0.7]{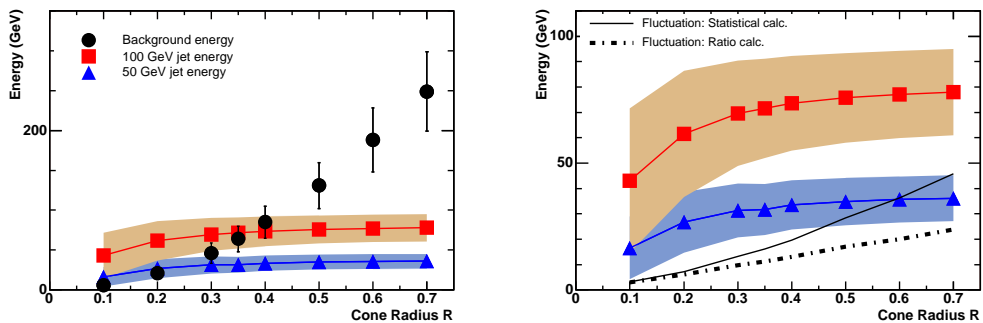}
\caption{\label{fig:BGJet}Left panel: Mean and RMS of background
  energy from central Pb+Pb background events (circles) compared to
  jet energy (simulated) from $p+p$ events for 50 GeV (triangles)
  and 100 GeV (squares) jets within jet cones of varying size $R$. The
  shaded bands around the jet energy symbols and vertical bars on the
  background symbols represent the RMS values of the distributions.
  Right panel: Jet energy within the cone for varying cone sizes
  (triangles: 50 GeV jets, squares: 100 GeV jets) compared to the
  central Pb+Pb background energy RMS in the cone calculated
  using two methods (solid line: statistical, dot-dashed line: ratio
  method). The bands represent the RMS of the jet energy
  distributions. }
\end{figure}

The left panel of Fig.~\ref{fig:BGJet} shows a comparison of the
jet energy and RMS for 50 GeV (triangles) and 100 GeV
jets (squares) in $p+p$ events and the total background energy
(circles) from uncorrelated particle production in Pb+Pb events, as
a function of cone radius $R$. All points include a $p_{T}$-cut of
2 GeV/c on charged tracks which rejects most of the background
from charged particles (98\% on average in central
{\sc hijing} events).  While the measured
jet energy only increases for small cone radii, up to $R\sim0.3$,
the background energy increases quadratically with $R$, exceeding
100 GeV at $R\sim0.4$. However, it is not the
magnitude of the background energy, but rather the event-by-event fluctuations
(represented by the vertical bars on the circle symbols in
Fig.~\ref{fig:BGJet}) which provide the challenge to jet reconstruction
and energy resolution.

The energy resolution for jets in heavy-ion events consists of two
components: the intrinsic energy resolution that would be achieved
without backgrounds and a contribution from the fluctuations of the
background energy contribution in the jet cone. The relative energy
resolution $\sigma(E_T)/E_T$ for full jet reconstruction in the
presence of backgrounds can be estimated as the quadratic sum of the
jet energy resolution $\sigma(E_T^{jet})$ in $p+p$ collisions and the
fluctuations of the background energy, \EBGtrue,
around the estimated background \EBGest:
\begin{equation}
  \sigma(E_T)/E_T  \lesssim \frac{1}{E_T}
  \sqrt{\sigma(E_T^{jet})^{2}+\sigma(\EBGtrue-\EBGest)^2}\label{eqn:res} 
\end{equation}

The right panel of Fig.~\ref{fig:BGJet} shows again the 
jet energy for 50 GeV (triangle markers) and 100 GeV jets (square
markers). The solid line shows the RMS of the background energy
(vertical bars in left panel) from central Pb+Pb events and thus
indicates the constribution of background fluctuations to the jet
energy resolution when using a simple average to estimate the level of background in the cone. 

One of the sources of background energy fluctuations is fluctuations
in the impact parameter of the collisions. The contribution of impact
parameter fluctuations to the final jet energy resolution can be
suppressed by estimating the energy from uncorrelated particles on an
event-by-event basis from the total energy deposited outside the jet
cone. For maximum statistical precision, we used the average total
$E_T$ per cell from the entire area of the jet-finding grid outside
the cone, without applying the \pT-cut on charged tracks,
$\langle E_T^{cell,nocut}\rangle$, as the basis for the background
energy estimate. The actual background energy inside the jet-cone,
with cuts, is then estimated by multiplying
$\langle E_T^{cell,nocut}\rangle$ by an average correction factor $F$
to account for the effect of the \pT-cut. The factor $F$ is
calculated as the ratio of the average cell-energy in the jet-finding
grid with cuts to the case without cuts, averaged over the entire sample of
background events. When the analysis is applied to experimental data,
$F$ can be calculated from events without detectable jets.

The resulting fluctuations of the true background energy (from {\sc
hijing} Pb+Pb events) around the estimated background energy using this 
procedure is indicated by the dashed line in the right panel
of Fig.~\ref{fig:BGJet}. The event-by-event
estimate of the background reduces the effect of fluctuations by about
a factor 2. As a result, larger cone radii can be used for
jet-finding. The remaining fluctuations are dominated by fluctuations
due to the finite statistical precision of the background estimate.

\subsection{Choice of parameters: cone radius}

\begin{figure}[!hbt]
  \center
  \includegraphics[scale=0.4]{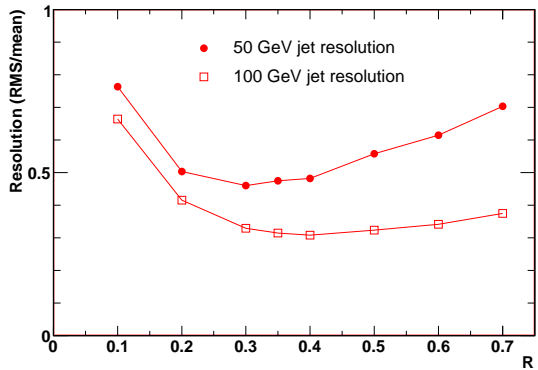}
  \caption{\label{fig:ResR} Energy resolution calculated
  as described in the text for 50~GeV jets (circles) and 100 GeV jets
  (squares) as a function of the cone radius $R$. }
\end{figure}

Figure~\ref{fig:ResR} shows the dependence of the relative jet energy
resolution on the cone radius $R$. The relative jet energy
resolution $\sigma(E_T)/E_T$ was calculated using Eq.~\ref{eqn:res}. The
RMS of the jet energy in $p+p$ collisions as shown in
Fig.~\ref{fig:BGJet} was used for $\sigma(E_T^{jet})$ and the RMS of the
background fluctuations (dash-dotted line in Fig.~\ref{fig:BGJet}) was
used for $\sigma(\EBGtrue-\EBGest)$.  For small values of $R$, the jet
energy resolution improves with increasing $R$ because
in-and-out-of-cone fluctuations dominate (first term in
Eq.~\ref{eqn:res}). At larger $R$, the background fluctuations
dominate and the resolution degrades with increasing $R$.  For this
study we chose the cone radius ($R=0.3$) which resulted in the best
energy resolution for 50 GeV jets. This result demonstrates that for optimum jet resolution in heavy-ion collisions, the jet
cone must be restricted in size, with an optimum size that decreases with decreasing
jet energy. Such restrictions will be necessary to enable jet reconstruction to the lowest
jet energies in central Pb+Pb collisions.  The biases introduced on the jet selection by
such restrictions will require extensive systematic study, including comparisons
with p+p reference data and theoretical calculations.


\subsection{Choice of parameters: seed energy and minimum jet energy}\label{sec:SeedOpt}

Finding the optimal values for the jet seed energy, $E_{T}^{seed}$, and
minimum accepted cone energy, $E_{T}^{cone}$, involves a trade-off between
jet-finding efficiency and sample purity.  To study this trade-off, we study
the results of {\sc hija} on `parameterized Pb+Pb' events. 
These events contain only pions and kaons sampled from the $p_{T}$ and pseudorapidity
distributions of the particles in {\sc hijing} events and are therefore
free of jets by construction. The parameterized Pb+Pb
events had a fixed charged particle rapidity density
of $dN_{ch}/dy = 4~000$ at mid-rapidity.

Using large values of $E_{T}^{seed}$ and/or $E_{T}^{cone}$ reduces the
number of `fake' jets (i.e. jets reconstructed by the algorithm which
are not the embedded {\sc pythia} jet) reconstructed in the
parameterized Pb+Pb events, but also reduces the number of
reconstructed jets in the event sample with embedded jets. 
The final parameters were selected through a trade-off between
fake rate and the jet-finding efficiency for 50 GeV jets.



\begin{table}[!hbt]
  \center \small
  \begin{tabular}{c|rl}\hline \hline
    \textbf{Parameter}
    &
    \multicolumn{2}{c}{
    \begin{tabular}{c}
    \textbf{Value}
    \end{tabular}}
    \\
    \hline
    $E_{T}^{seed}$
    &
    4.6
    &
    GeV
    \\
    $R$
    &
    0.3
    &
    \\
    $E_{T}^{cone}$
    &
    14.0
    &
    GeV
    \\
    $p_{T}^{cut}$
    &
    2.0
    &
    GeV/$c$
    \\
    \hline \hline
\end{tabular}
\protect\caption{Values of algorithm parameters after optimisation. }
\protect\label{tab:ParamFinal}
\end{table}

\section{Results}
This section summarizes the {\sc hija} algorithm results for jet
efficiency, energy and direction resolution in simulated Pb+Pb
collisions.

\subsection{Jet-finding efficiency and direction resolution}

The values for $E_{T}^{seed} =
4.6$ GeV and $E_{T}^{cone} = 14.0$ GeV were selected, as shown in  Table~\ref{tab:ParamFinal}, because they
resulted in (1) a high efficiency for finding 50 GeV embedded jets in
Pb+Pb events (greater than 70$\%$) and (2) a low number of `fake' jets
reconstructed in simulated Pb+Pb events (a rate of about 3$\%$ in
parameterized {\sc hijing} compared to about one signal jet expected
per central Pb+Pb collision).
The fraction of events accepted as containing jets for each type
of event is shown in Table~\ref{tab:exclusion}. As discussed in
Sect.~\ref{sec:SeedOpt}, parameterized {\sc hijing} events do not contain jets by construction.
The true fake rate is expected to lie between the parameterized and pure {\sc hijing}
values since pure hijing events do contain jets which are not able to be tagged in
the simulations.

The final values of all the optimised algorithm parameters used in this study are
presented in Table~\ref{tab:ParamFinal}.


\begin{table}[!hbt]
  \center \small
  \begin{tabular}{ccccc}\hline \hline

    &
    \begin{tabular}{c}
      \textbf{Param.} \\\textbf{{\sc hijing}} \\
    \end{tabular}
    &
    \begin{tabular}{c}
      \textbf{Pure} \\ \textbf{{\sc hijing}} \\
    \end{tabular}
    &
    \begin{tabular}{c}
    \textbf{50 GeV jets} \\\textbf{+ {\sc hijing}}\\
    \end{tabular}
    &
    \begin{tabular}{c}
    \textbf{100 GeV jets} \\\textbf{+ {\sc hijing}}\\
    \end{tabular}
    \\
    \hline
    \begin{tabular}{l}
      \textbf{Accepted}\\
    \end{tabular}
    &
    $3\%$
    &
    $13\%$
    &
    $70\%$
    &
    $97\%$
    \\
    \hline \hline
\end{tabular}
\protect\caption{Percentage of the event sample accepted by the algorithm as containing a jet using
  $E_{T}^{seed}=4.6$ GeV and $E_{T}^{cone}=14.0$ GeV. }
\protect\label{tab:exclusion}
\end{table}

\begin{figure}[!hbt]
  \center
  \includegraphics[scale=0.4]{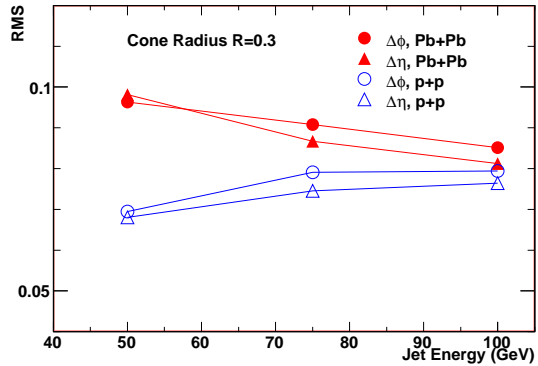}
  \caption{\label{fig:DirRes} RMS of jet $\Delta \eta$ and $\Delta
  \phi$ distributions for the Pb+Pb (solid symbols) and $p+p$ (open
  symbols) cases as a function of jet energy for $R=0.3$. }
\end{figure}

The high accuracy with which {\sc hija} reconstructs jet directions is
shown in Fig.~\ref{fig:DirRes} by the RMS values of the difference
between the reconstructed and input jet directions (calculated by {\sc pycell}), $\Delta
\eta$ (triangles) and $\Delta \phi$ (circles) for Pb+Pb (solid) and
$p+p$ (open).  The high background in Pb+Pb collisions affects the
direction resolution in $\eta$ and $\phi$ similarly,
leading to almost equal resolutions in both directions.

For the Pb+Pb case, the algorithm becomes more accurate with
increasing jet energy because the signal to background ratio
increases.  For $p+p$ on the other hand, the jet
direction resolution becomes slightly worse at higher jet energy.
This is due to the small cone radius of $R=0.3$, which occasionally
leads to two reconstructed jets instead of one.  This ``splitting''
effect results in a small fraction of large $\Delta \eta$ and $\Delta
\phi$ values, increasing with jet energy. 
The effect is also present in Pb+Pb, but it is offset
by the background fluctuations. Possible corrections for this effect are
discussed in ~\cite{RunII}.

\subsection{Jet energy resolution}

\begin{figure}[!hbt]
  \center
  \includegraphics[width=\textwidth]{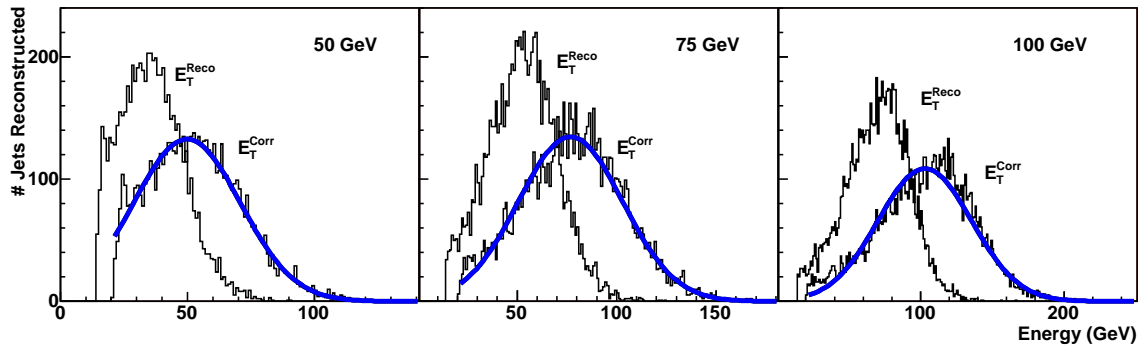}
  \caption{\label{fig:EShift} Reconstructed ($E^{Reco}_{T}$) and corrected ($E^{Corr}_{T}$)
        jet energy distributions for 50 GeV, 75 GeV and 100 GeV jets
        embedded in central Pb+Pb {\sc hijing} events.}
\end{figure}

Figure~\ref{fig:EShift} shows the energy distributions of
reconstructed jets that have been embedded in simulated central Pb+Pb events, using the optimized
algorithm parameters (Table~\ref{tab:ParamFinal}). Distributions are
shown for both the raw reconstructed jet energy $E^{Reco}_T$ and the
corrected jet energy $E^{Corr}_{T}$.  The reconstructed jet energies
$E^{Reco}_{T}$ were corrected for losses due to the small cone
radius, the track $p_{T}$ cut, and missing energy from unmeasured
particles by multiplication with an average correction factor
$C=1/0.6731$, calculated from the
({\sc pythia}) simulations. The factor is averaged with cross-sectional
weight, and therefore reproduces the actual jet energy better at low energy.
The solid lines
represent Gaussian fits to the corrected jet energy distributions
and are used to extract the width $\sigma$ of the distributions.

The mean values and $\sigma$ of the corrected energy distributions
are given in Table~\ref{tab:EShift}. The mean values
are within $4\%$ of the input jet energies for all three samples.

\begin{table}[!hbt]
  \center \small
  \begin{tabular}{cccc}\hline \hline

    &
    \begin{tabular}{c}
      \textbf{50 GeV jets}\\ \textbf{+ {\sc hijing}}\\
    \end{tabular}
    &
    \begin{tabular}{c}
    \textbf{75 GeV jets}\\ \textbf{+ {\sc hijing}}\\
    \end{tabular}
    &
    \begin{tabular}{c}
    \textbf{100 GeV jets} \\ \textbf{+ {\sc hijing}}\\
    \end{tabular}
    \\
    \hline
    \begin{tabular}{l}
      \textbf{$\langle E^{Reco}_{T} \rangle \pm \sigma$}  \textbf{(GeV)} \\
    \end{tabular}
    &
    $34 \pm 14$
    &
    $52 \pm 18$
    &
    $70 \pm 22$
    \\

    \begin{tabular}{l}
      \textbf{$\langle E^{Corr}_{T} \rangle \pm \sigma$} \textbf{(GeV)} \\
    \end{tabular}
    &
    $50 \pm 21$
    &
    $77 \pm 26$
    &
    $103 \pm 33$
    \\
    \hline \hline
\end{tabular}
\protect\caption{Mean value and standard deviation ($\sigma$) (taken from the Gaussian fits) of
the reconstructed jet energy distributions (embedded in central Pb+Pb {\sc hijing} events) for various input
jet energies before and after correction for losses due to the small
cone radius, the track $p_{T}$ cut, and missing energy
from unmeasured particles.}
\protect\label{tab:EShift}
\end{table}

\section{Comparison to $p+p$ collision baseline}
To further separate the effects of the various algorithm cuts and
the effect of background fluctuations, we have studied the effects of
the algorithm cuts on $p+p$ simulations. For this purpose a cross-section
weighted spectrum of {\sc pythia} jets with {\sc pycell} transverse energies
between 20 and 180 GeV was generated. The influence of the 
{\sc hija} cuts was examined by performing jet-finding directly on the {\sc pythia}
particle lists without detector simulation. {\sc hija} results for jet-finding using all particle
information and cone radius $R=1.0$, agree with the reference
spectrum from {\sc pycell}.

\begin{figure}[!hbt]
  \center
  \includegraphics[scale=0.4]{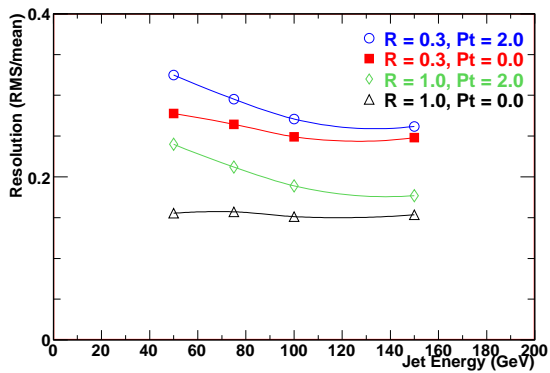}
  \caption{\label{fig:NoRes} Energy resolution as a function of
  input jet energy for various {\sc hija} cut combinations on p+p {\sc pythia}
  events without detector simulation.}
\end{figure}

Figure~\ref{fig:NoRes} shows the resolutions using {\sc hija} on the {\sc
pythia} spectrum, as a function of energy, for two choices of $R$
and $p_T$-cut. To determine the energy resolution as a function of
jet energy, jets were selected from the generated spectrum in a narrow
($\pm 5$ GeV) energy range around the values indicated by the data points. The
spread of the reconstructed energies due to the width of the selected
interval was taken out by subtracting the nominal trend of
reconstructed energy as a function of input energy. The resolution for
the ideal case (no cuts), but excluding undetectable particles ($\nu$,
$K_{L}$ and neutrons) is approximately 15$\%$, independent of energy
(triangle symbols). Application of the cuts on charged particle $p_T$
and reducing the cone radius to $R=0.3$ leads to additional loss of
resolution as shown by the diamond and square markers. Using the
final heavy-ion optimized parameters (see Table~\ref{tab:ParamFinal})
leads to a resolution of approximately 26-35$\%$ for jets with 50
GeV~$< E_{T} <$~160 GeV (circles). This is the intrinsic limit of the
jet resolution in $p+p$ using {\sc hija} with cuts optimized for central Pb+Pb.

\begin{figure}[!hbt]
  \center
  \includegraphics[scale=0.4]{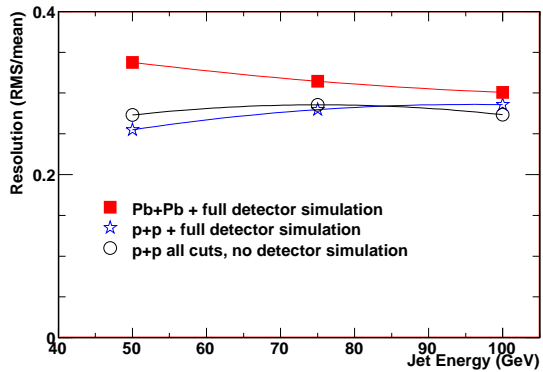}
  \caption{\label{fig:ERes}Reconstructed jet resolution (RMS/mean) as a function of energy for
  the Pb+Pb case including detector effects (solid squares), $p+p$ case including detector
effects (open stars) and $p+p$ case without detector effects (open
  circles). }
\end{figure}

Figure~\ref{fig:ERes} illustrates the effect of detector energy
resolution and background fluctuations in Pb+Pb events by comparing
the `ideal' jet energy resolution in $p+p$
({\sc pythia} without detector simulation, open circles) with the
actual jet energy resolution obtained with {\sc hija} including all detector
effects for the $p+p$ (open stars) case and detector and high
multiplicity background effects for the Pb+Pb (solid squares) case. For
this figure, the $E_T^{cone}$ cut was applied consistently in all cases
to enable a direct comparison.  The addition of the $E_T^{cone}$ cut
artificially improves the resolution by a small amount compared to
Figs.~\ref{fig:ResR} and \ref{fig:NoRes} as it truncates the
distribution at low reconstructed $E_{T}$, leading to a reduced RMS
value.

Detector effects (open stars) do not produce significant
differences compared to the pure {\sc pythia} case (open circles).
The small loss in jet energy resolution in $p+p$ from 50 to 100 GeV is
an artifact of the $E_T^{cone}$ cut.  In Pb+Pb the contribution from the 
fluctuating background leads to an additional spread of the
reconstructed jet energy, varying from $\sim$8$\%$ for 50 GeV jets, 
down to $<$2$\%$ for 100 GeV jets.

\section{Summary and Conclusions}

A UA1-based cone algorithm has been adapted to reconstruct jets in
Pb+Pb collisions at $\sqrt{s_{NN}}$ = 5.5 TeV at the LHC. A technique
to estimate and subtract the background in heavy-ion collisions on an
event-by-event basis has been developed. The contributions to the jet
energy resolution from in-and-out-of-cone fluctuations, undetectable
particles and the various algorithm cuts were studied. It has been
shown that using this algorithm with the ALICE detectors, jets of 50
GeV and higher transverse energies can be reconstructed on an event-by-event
basis. The $p+p$ resolutions are significantly affected by the choice
of parameters required to suppress the background in heavy-ion collisions. 
The final resolutions obtained with the selected algorithm parameters is
$\sim$34$\%$($\sim$26$\%$) for 50 GeV jets and
$\sim$30$\%$($\sim$28$\%$) for 100 GeV jets in Pb+Pb($p+p$)
collisions. The main contribution to the degradation of the jet energy
resolution in Pb+Pb compared to $p+p$ is due to the fluctuating
underlying event.  This background is intrinsic to heavy-ion
collisions and will be present in all experiments at the LHC.  Using
these techniques, the modifications to intermediate energy partons
interacting in the dense nuclear medium at LHC should be
experimentally accessible, providing new insights into the color
structure of the quark-gluon plasma.

\section*{Acknowlegements}
SLB and HG would like to acknowledge the financial support of the
National Research Foundation, South Africa. 
ORNL is managed by UT-Battelle, LLC, for the U.S. Department of Energy
under contract DE-AC05-00OR22725. LBNL 
was supported by the Office of Science, Nuclear Physics,
U.S. Department of Energy under Contract No. DE-AC03-76SF00098.

\end{document}